\documentclass[3p,twocolumn]{elsarticle}

\usepackage{amsfonts,amsmath,amssymb}
\usepackage{graphicx,color}

\def\blfootnote{\xdef\@thefnmark{}\@footnotetext}

\begin{document}

\begin{frontmatter}

\title{Statistical Production of $B_c$ Mesons in Heavy-Ion Collisions at the LHC Energy}

\author[a]{Shouxing Zhao}
\author[a]{Min He}

\address[a]{Department of Applied Physics, Nanjing University of Science and Technology, Nanjing 210094, China}

\date{\today}

\begin{abstract}
The recombination production of $B_c$ mesons in heavy-ion collisions at the LHC energy is facilitated by the abundant and highly thermalized charm ($c$) quarks transported in the deconfined medium created. We study the production of $B_c$ mesons via $c$ and bottom ($b$) quark recombination in a statistical fashion by placing $B_c$ in the position of a member of the family of open $b$ hadrons, which allows us to make quantitative predictions for the modifications of the production fraction ($f_c$) of $B_c$ mesons and its relative production to $B$ mesons in $\sqrt{s_{\rm NN}}=5.02$\,TeV Pb-Pb collisions with respect to proton-proton ($pp$) collisions at the same energy. The statistical production yield of $B_c$ mesons is converted into the transverse momentum ($p_T$) distribution with the shape computed from resonance recombination using the $c$- and $b$-quark phase space distributions that have been simulated via Langevin diffusion and constrained by open $c$- and $b$-hadron observables. Supplemented with the component fragmented from $b$-quark spectrum that dominates at high $p_T$, the total $p_T$ spectrum of $B_c$ mesons is obtained and converted into the $p_T$ dependent nuclear modification factor ($R_{\rm AA}$). Both $f_c$ and the integrated $R_{\rm AA}$ exhibit a $\sim5$-fold enhancement in central Pb-Pb collisions relative to the $pp$ reference. Comparison with data measured by the CMS experiment shows decent agreement within theoretical and experimental uncertainties.

\end{abstract}


\end{frontmatter}

\section{Introduction
\label{sec_intro}}

The $B_c$ mesons as bound states of a bottom ($b$) quark with an anticharm ($\bar{c}$) quark, or vice versa, provide a unique avenue to the understanding of heavy quark dynamics. They are intermediate between charmonium ($c\bar{c}$) and bottomonium ($b\bar{b}$) states in terms of mass, size and binding energy~\cite{Eichten:1994gt,Ebert:2002pp,Godfrey:2004ya,Eichten:2019gig}. Being composed of two different heavy flavors, $B_c$ mesons cannot annihilate into gluons; consequently excited $B_c$ states lying below the open $BD$ threshold can only undergo radiative or hadronic transitions into the pseudoscalar ($^1S_0$) ground state $B_c$ that decays weakly, leading to total widths less than a few hundred keV~\cite{Eichten:1994gt,Ebert:2002pp}, significantly smaller than their charmonia and bottomnia counterparts~\cite{ParticleDataGroup:2022pth}.

The production of $B_c$ mesons in hadronic ({\it e.g.} proton-proton, $pp$) collisions entails the simultaneous creation of a $c\bar{c}$ and a $b\bar{b}$ pair in a single collision, rendering it much rarer than that of other mesons containing a single $b$ quark. In $pp$ collisions at the LHC energy, the LHCb experiment has measured the production fraction of the ground state $B_c^-$ to be $\sim0.26$\% (with significant uncertainty due to the one from the branching fraction of the decay $B_c^-\rightarrow J/\psi\mu^-\bar{\nu}$ used in the measurement) relative to the total $b\bar{b}$ cross section~\cite{LHCb:2019tea}, which is negligibly small compared to the fractions of $B$ mesons and $\Lambda_b$ baryons~\cite{HFLAV:2019otj,He:2022tod}.

However, in relativistic heavy-ion collisions where a deconfined medium (Quark-Gluon Plasma, QGP) is created, abundant heavy quarks are produced through primordial hard processes and then transported in the QGP~\cite{Rapp:2018qla,Dong:2019byy,Dong:2019unq,Apolinario:2022vzg,He:2022ywp}, leading to a new production mechanism for $B_c$ mesons, namely the recombination of $b$ and $c$ quarks from different primordial nucleon-nucleon collisions that could significantly enhance the $B_c$ yield~\cite{Schroedter:2000ek}. For example, in the most central 0-10\% Pb-Pb collisions at the LHC energies, $dN_{c\bar{c}}/dy\sim 20$~\cite{He:2019vgs} charm and $dN_{b\bar{b}}/dy\sim 0.9$~\cite{He:2022tod} bottom quarks per unit rapidity can be produced at mid-rapidity, respectively. In events in which there is a single $b\bar{b}$ pair along with a number of $c\bar{c}$ pairs, the $b$ quark can find any of the $c$ quarks near in phase space to recombine and produce a $B_c$ bound state. This recombination process is particularly strengthened given the high degree of thermalization of $c$ quarks in the QGP~\cite{He:2022ywp,He:2019vgs,He:2011yi} as primarily evidenced by the large elliptic flow of $D$ mesons~\cite{ALICE:2021rxa} measured in Pb-Pb collisions at the LHC energy.

This novel production mechanism makes $B_c$ an attractive part of the endeavor of using heavy quarkonia as a probe of the QGP properties~\cite{Rapp:2008tf,Braun-Munzinger:2009dzl,Rothkopf:2019ipj,Chapon:2020heu,Yao:2021lus,Zhao:2020jqu,Andronic:2024oxz}. In this context, the recombination production of $B_c$ mesons, since it was first proposed~\cite{Schroedter:2000ek}, has been studied in various transport~\cite{Liu:2012tn,Zhao:2022auq,Wu:2023djn} and instantaneous coalescence~\cite{Chen:2021uar} models. The static in-medium properties of $B_c$ mesons such as binding energy and radius used as inputs of these models are sensitive to the employed heavy quark potential that is currently under hot debate~\cite{Liu:2015ypa,Bazavov:2023dci}. On the experimental side, pioneering measurement by the CMS experiment in $\sqrt{s_{\rm NN}}=5.02$\,TeV Pb-Pb collisions, although restricted to relatively large transverse momenta ($p_T>6$\,GeV), indeed gives a first hint that the $B_c$ production in the presence of QGP is enhanced relative to $pp$ collisions, as indicated by the nuclear modification factor well above unity in the lower $p_T$ bin accessed in the measurement~\cite{CMS:2022sxl}. One notes that, although the recombination of $B_c$ depends linearly on the $c$ quark densities, weaker than the quadratic dependence in the case of charmonia regeneration~\cite{Wu:2024gil}, the nuclear modification factor of $B_c$ is likely to to be significantly larger than that of $J/\psi$ mesons~\cite{He:2021zej}, owing to the too small production of the former in $pp$ collisions as the reference.

In this work, we treat $B_c$ as a member of the family of $b$-hadrons that contain a single $b$ quark and study its production in $\sqrt{s_{\rm NN}}=5.02$\,TeV Pb-Pb collisions in the statistical hadronization model (SHM) following the spirit of the statistical production of charmonia and $c$-hadrons~\cite{Braun-Munzinger:2000csl,Andronic:2006ky,Andronic:2021erx,Andronic:2023tui}. For heavy quarkonia that feature deep binding, statistical coalescence (recombination) does not apply to their production in elementary collisions, in contrast to the case of heavy-light hadrons where SHM empirically works well~\cite{He:2022tod,Dai:2024vjy}. However, the high degree of thermalization for $c$ quarks facilitates the statistical description of not only charmonia~\cite{Braun-Munzinger:2000csl,Gorenstein:2000ck} but also $B_c$ mesons in Pb-Pb collisions at the LHC energy. For the latter, one can envisage that a single $b$ quark is surrounded by a number of highly thermalized diffusing $c$ quarks and stochastic (statistical) recombination satisfying phase space conditions occurs efficiently, which is reminiscent of the case of statistical $c$-light quark recombination to form open $c$-hadrons~\cite{Andronic:2021erx}. Given the small number of heavy quark production, it turns out important to adopt the canonical ensemble version of SHM to exactly conserve the $b$ and $c$ numbers for predicting the absolute yield of $B_c$ mesons. This statistical production description in particular allows us to make quantitative predictions for the production fraction of $B_c$ mesons relative to the total $b\bar{b}$ multiplicity and the enhancement of its relative production to $B$ mesons in the presence of QGP with respect to $pp$ collisions.

This article is organized as follows. In Sec.~\ref{sec_Bc-pp}, we fix $B_c$'s $p_T$ differential cross section in $\sqrt{s}=5.02$\,TeV $pp$ collisions by making use of the previously computed $B^-$'s differential cross section in combination with the $p_T$ dependent $B_c^-/B^-$ ratio measured by LHCb experiment. This will be used as the reference for the ensuing calculation of $B_c$'s nuclear modification factors and also enables us to make an estimate of its integrated cross section. In Sec.~\ref{sec_Bc-SHM}, we elaborate on the SHM calculation of $B_c$ production in $\sqrt{s_{\rm NN}}=5.02$\,TeV Pb-Pb collisions, putting it in the context of the statistical production of the whole family of $b$-hadrons such that the $b$-number conservation through hadronization is taken into full account and serves as an constraint for the prediction of the production fraction of $B_c$ mesons. In Sec.~\ref{sec_Bc-pT-spectrum-RAA}, we compute the $B_c$'s $p_T$ distribution by normalizing its $p_T$ spectrum from resonance recombination of realistically transported $c$ and $b$ quarks to its statistical production yield, which is then supplemented by the component from fragmentation of $b$ quarks that dominates at high $p_T$. The total $p_T$ spectrum is converted into the $p_T$ dependent nuclear modification factors and compared to CMS data. We conclude in Sec.~\ref{sec_summary}.

\section{$B_c$'s $p_T$ differential cross section in $pp$ collisions}
\label{sec_Bc-pp}
In this section, we construct the $p_T$ differential cross section $d\sigma/dp_Tdy$ for the ground state $B_c^-$ in $\sqrt{s}=5.02$\,TeV $pp$ collisions at mid-rapidity as the reference to measure its $p_T$ spectral modifications in Pb-Pb collisions. The production fraction of the ground state pseudoscalar $B_c^-$ mesons relative to the sum of those of $B^-$ and $\bar{B}^0$ ($f_c/(f_u+f_d)$) has been measured by the LHCb experiment in $\sqrt{s}=7$ and 13\,TeV $pp$ collisions in the kinematic regions of transverse momentum $4<p_T<25$\,GeV and pseudo-rapidity $2.5<\eta<4.5$, showing almost the same results between these two energies~\cite{LHCb:2019tea}. The measured ratio shows no rapidity dependence and its rather weak $p_T$ dependence has been linearly fitted~\cite{LHCb:2019tea}, which is converted into $B_c^-/B^-$ assuming isospin symmetry ($f_u=f_d$) and plotted in Fig.~\ref{fig_Bc-vs-B-PbPb}(b) (blue band), where the significant spread is due to the uncertainty from the branching fraction of $\mathcal{B}(B_c^-\rightarrow J/\psi\mu^-\bar{\nu})=1.95\%\pm 0.46\%$~\cite{LHCb:2019tea} used in the measurement.

\begin{figure} [!tb]
\vspace{-0.5cm}
\includegraphics[width=1.05\columnwidth]{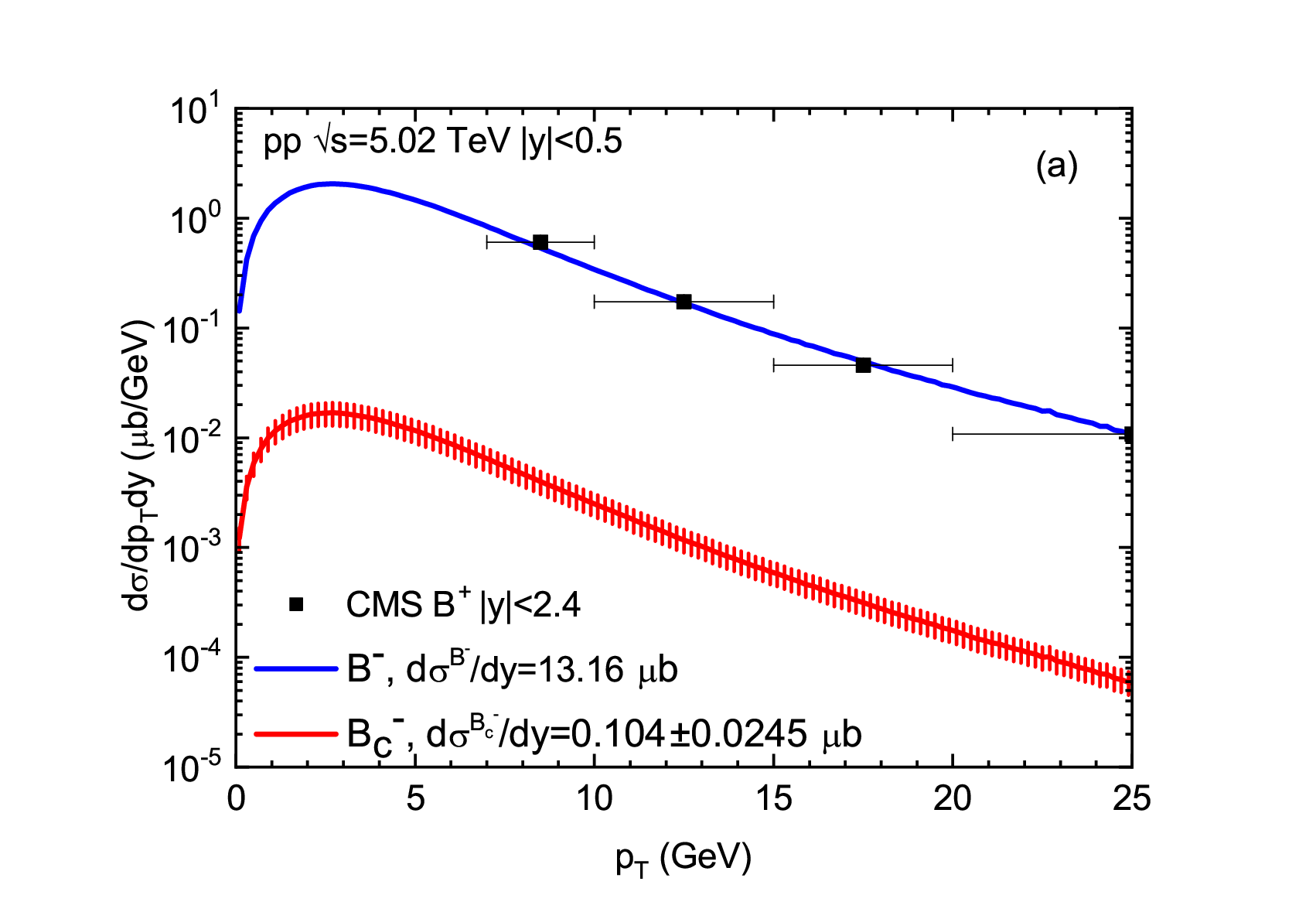}
\vspace{-0.5cm}
\includegraphics[width=1.05\columnwidth]{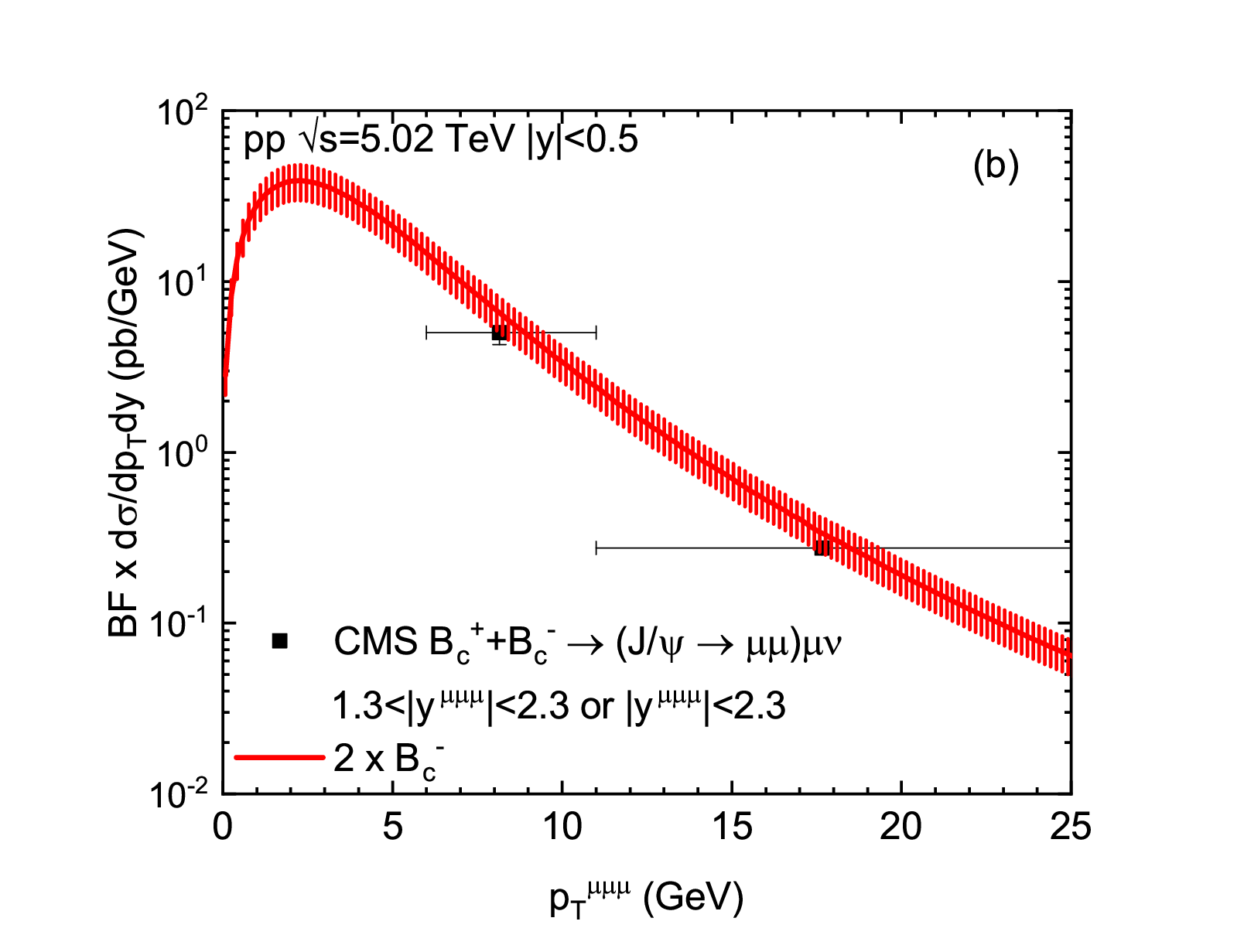}
\vspace{-0.5cm}
\caption{(a) The constructed $p_T$ differential cross section for the ground state $B_c^-$ and that for $B^-$ quoted from~\cite{He:2022tod} in comparison with CMS data~\cite{CMS:2017uoy} at mid-rapidity in $\sqrt{s}=5.02$\,TeV $pp$ collisions. (b) The computed $B_c^-+B_c^+$ differential cross section multiplied by pertinent branching fractions (BF) as a function of $p_T^{\mu\mu\mu}$ in comparison with CMS data~\cite{CMS:2022sxl}. See text for more details.}
\label{fig_Bc-pT-spectrum-pp}
\end{figure}

Assuming the $p_T$-dependent $B_c^-/B^-$ to remain unchanged, we construct $B_c^-$'s $d\sigma/dp_Tdy$ at mid-rapidity in $\sqrt{s}=5.02$\,TeV $pp$ collisions by multiplying the $B_c^-/B^-$ with $B^-$'s $d\sigma/dp_Tdy$ that has been determined in~\cite{He:2022tod}. The resulting $d\sigma/dp_Tdy$ for $B_c^-$ and that for $B^-$ quoted from~\cite{He:2022tod} (in comparison with CMS measurement~\cite{CMS:2017uoy}) are displayed in Fig.~\ref{fig_Bc-pT-spectrum-pp}(a). As a byproduct, we've estimated the $B_c$'s integrated cross section to be $d\sigma^{B_c^-}/dy=0.104\pm0.0245~\mu b$ at mid-rapidity and its relative ratio to $d\sigma^{B^-}/dy=13.16~\mu b$ for $B^-$ to be $B_c^-/B^-=0.00790\pm0.00186$. Based on this ratio, the production fraction of $B_c^-$ was  also estimated to be $f_c=0.00258\pm0.00062$~\cite{LHCb:2019tea} (only the dominant uncertainty from $\mathcal{B}(B_c^-\rightarrow J/\psi\mu^-\bar{\nu})$ is quoted).

The CMS experiment has measured the $d\sigma/dp_Tdy$ for $B_c^-+B_c^+$ multiplied by the pertinent branching fractions through a two-step decay process $B_c\rightarrow (J/\psi\rightarrow\mu\mu)\mu\nu$ in two bins of the three-muons' transverse momentum $6<p_T^{\mu\mu\mu}<11$\,GeV at $1.3<|y^{\mu\mu\mu}|<2.3$ and $11<p_T^{\mu\mu\mu}<35$\,GeV at $|y^{\mu\mu\mu}|<2.3$~\cite{CMS:2022sxl}. To compare to the CMS data, we multiply the $B_c^-$'s $d\sigma/dp_Tdy$ constructed above by $\mathcal{B}(B_c^-\rightarrow J/\psi\mu^-\bar{\nu})=1.95\%\pm 0.46\%$~\cite{LHCb:2019tea} and $\mathcal{B}(J/\psi\rightarrow\mu\mu)=5.96\%$ (with negligible uncertainty)~\cite{ParticleDataGroup:2022pth}. Taking into account the average $p_T$ shift of $p_T^{\mu\mu\mu}=0.85\cdot p_T^{B_c}$~\cite{CMS:2022sxl}, the result (the factor $2$ is to account for the sum of $B_c^-+B_c^+$) is plotted in Fig.~\ref{fig_Bc-pT-spectrum-pp}(b) in comparison with the CMS data~\cite{CMS:2022sxl}. The central values of CMS data in two rather wide $p_T$ bins are better described by using the lower bound of $\mathcal{B}(B_c^-\rightarrow J/\psi\mu^-\bar{\nu})$.

\section{Statistical production of $B_c$ mesons in Pb-Pb collisions}
\label{sec_Bc-SHM}
In this section, we compute the {\it absolute} production yields of $B_c^-$ in different centralities of $\sqrt{s_{\rm NN}}=5.02$\,TeV Pb-Pb collisions from the SHM. For large systems like heavy-ion collisions, quantum charges such as electric charge ($Q$), baryon number ($N$) and strangeness ($S$) can be treated grand-canonically in the statistical description. In contrast, the small number of $c$-hadrons ($C$) and $b$-hadrons ($B$) renders their {\it exact} conservation through the canonical ensemble description essential, such that the partition function reads~\cite{Dai:2024vjy,Becattini:2009sc}
\begin{align}\label{Z-CB}
Z(C,B)=\lambda_{Q}\lambda_{N}\lambda_{S}\frac{1}{(2\pi)^2}\int_{0}^{2\pi} d\phi_{C} d\phi_{B} e^{i (C\phi_{C}+B\phi_{B})} \nonumber\\
\times{\rm exp} [\sum_j \gamma_s^{N_{sj}} \gamma_c^{N_{cj}} \gamma_b^{N_{bj}} e^{-i(C_j\phi_{C}+B_j\phi_{B})}z_j],
\end{align}
where $\lambda_Q$, $\lambda_N$ and $\lambda_S$ are fugacities associated with $Q$, $N$ and $S$, respectively; $C_j$ and $B_j$ are the charm and bottom number of the $j$-th particle, respectively; $\gamma_s$, $\gamma_c$ and $\gamma_b$ are fugacities that account for the deviation from chemical equilibrium for hadrons containing $N_{sj}$, $N_{cj}$ and $N_{bj}$ strange, $c$ and $b$ quarks plus antiquarks, respectively. In Eq.~(\ref{Z-CB}), $z_j$ denotes the one-particle partition function
\begin{align}\label{1-partition-function}
z_j=(2J_j+1)\frac{VT_H}{2\pi^2}m_j^2K_2(\frac{m_j}{T_H}),
\end{align}
which also specifies the chemical equilibrium multiplicity of the $j$-th hadron of mass $m_j$ and spin $J_j$ in a fireball of volume $V$ under the Boltzmann approximation at hadronization temperature $T_H$, with $K_2$ being the modified Bessel function of the second kind. The primary multiplicity of a heavy hadron produced from the SHM is then given by
\begin{align}\label{SHM-yield}
\langle N_j\rangle = \gamma_s^{N_{sj}} \gamma_c^{N_{cj}} \gamma_b^{N_{bj}} z_j \frac{Z(C-C_j,B-B_j)}{Z(C,B)},
\end{align}
where $Z(C-C_j,B-B_j)/Z(C,B)$ (known as chemical factor less than unity for $c$ or $b$ hadrons in a neutral system with $C=B=0$~\cite{Dai:2024vjy}) characterizes the {\it canonical suppression} relative to the grand-canonical statistical production as a result of the {\it exact} conservation of $C$ and $B$.

To evaluate the partition function, hadrons to be summed over in the exponential of Eq.~(\ref{Z-CB}) are divided into five categories: light hadrons made of up, down and strange quarks with $C_j=B_j=0$, $N_{cj}=N_{bj}=0$ and $N_{sj}=0,1,2,3$; open $c$ hadrons with $C_j=\pm1$, $N_{cj}=1$, $B_j=0$,  $N_{bj}=0$ and $N_{sj}=0,1,2$; charmonia with $C_j=0$, $N_{cj}=2$, $B_j=0$, $N_{bj}=0$ and $N_{sj}=0$; open $b$ hadrons with $C_j=0$, $N_{cj}=0$, $B_j=\pm1$, $N_{bj}=1$ and $N_{sj}=0,1,2$; $B_c$ mesons with $C_j=\pm1$, $N_{cj}=1$, $B_j=\pm1$, $N_{bj}=1$ and $N_{sj}=0$; bottomonia with $C_j=0$, $N_{cj}=0$, $B_j=0$, $N_{bj}=2$ and $N_{sj}=0$. We have safely neglected heavy baryons that contain two $c$ or $b$ quarks since their masses are much larger than other open $c$ or $b$ hadrons. Clearly the summations over light hadrons, charmonia and bottomonia factorize out of the exponential and for the purpose of computing the heavy hadrons' multiplicity, the evaluation of the partition function reduces to
\begin{align}\label{Z_CB-reduced}
&Z(C,B)\propto \int_{0}^{2\pi} d\phi_{C} d\phi_{B}\cos(C\phi_{C}+B\phi_{B}) \nonumber\\
&\times{\rm exp}[2\sum_{j\in H^+} \gamma_s^{N_{sj}} \gamma_c^{N_{cj}} \gamma_b^{N_{bj}} \cos(C_j\phi_{C}+B_j\phi_{B})z_j].
\end{align}
Now the summation in the exponential of Eq.~(\ref{Z_CB-reduced}) is restricted to be over {\it positively} charged $c$ and $b$ hadrons, such as $D^0$, $D_s^+$, $\Lambda_c^+$, $B^0$, $B_s^0$, $\bar{\Lambda}_b^0$ and $B_c^+$. In Eqs.~(\ref{SHM-yield}) and~(\ref{Z_CB-reduced}), the strangeness fugacity is taken to be $\gamma_s=1$ ({\it i.e.}, strangeness chemical equilibrium is reached down to the hadronization in Pb-Pb collisions). But the $c$ and $b$ fugacities $\gamma_c$ and $\gamma_b$ should be self-consistently determined from the balance equations that stipulate $c$ and $b$ conservation through hadronization:
\begin{align}\label{c-b-balance-eq}
\frac{dN_{c\bar{c}}}{dy}=\sum_{j\in oc^+}\langle N_j\rangle + \sum_{j\in hc}\langle N_j\rangle + \sum_{j\in B_c^+}\langle N_j\rangle,\\
\frac{dN_{b\bar{b}}}{dy}=\sum_{j\in ob^+}\langle N_j\rangle + \sum_{j\in hb}\langle N_j\rangle + \sum_{j\in B_c^+}\langle N_j\rangle,
\end{align}
where the summations run over the primary multiplicities $\langle N_j\rangle$ (cf. Eq.~(\ref{SHM-yield})) of {\it positively} charged open $c$/$b$ hadrons ($oc^+$ and $ob^+$, respectively), charmonia/bottomonia ($hc$ and $hb$, respectively) and $B_c^+$ mesons. Since the majority ($\sim 99$\%) of $c$ content is carried by open $c$ hadrons, the $c$ fugacity $\gamma_c$ is essentially governed by the first term on the right hand side of Eq.~(\ref{c-b-balance-eq}); the same is true for $\gamma_b$.

\begin{table*}[!t]
\begin{center}
\begin{tabular}{lcccccc}
\hline\noalign{\smallskip}
$~$          & 0-20\%     & 20-40\%         & 40-60\%         & 60-80\%\\
\noalign{\smallskip}\hline\noalign{\smallskip}
$V_{\Delta y=1}({\rm fm^3})$ & 4170 & 1849 & 709 & 200 \\
$dN_{c\bar{c}}/dy$  &	17.2 & 6.46 & 2.15 & 0.44 \\
$dN_{b\bar{b}}/dy$  &	0.74 & 0.272 & 0.0806 & 0.0165 \\
\noalign{\smallskip}\hline\noalign{\smallskip}
$\gamma_c$         & 13.35         &  11.58          & 10.87         &  11.84 \\
$\gamma_b$         & $6.27\cdot10^7$ &  $7.59\cdot10^7$  & $1.03\cdot10^8$ & $1.62\cdot10^8$ \\

\noalign{\smallskip}\hline
\end{tabular}
\end{center}
\caption{Volume of the fireball, $c$- and $b$-quark multiplicities per unit rapidity at mid-rapidity and their fugacities in different centralities of $\sqrt{s_{\rm NN}}=5.02$\,TeV Pb-Pb collisions. For 0-20\% and 20-40\% centrality, 20\% reduction is applied to the $c$-quark multiplicity owing to shadowing effect. For the other two centralities, 10\% shadowing is applied.}
\label{tab_gamma_c-b}
\end{table*}

\begin{table*}[!t]
\begin{center}
\begin{tabular}{lcccccc}
\hline\noalign{\smallskip}
$dN/dy$          & 0-20\%     & 20-40\%         & 40-60\%         & 60-80\%\\

\noalign{\smallskip}\hline\noalign{\smallskip}
$B^-(=\bar{B}^0)$         & 0.23234 &  0.085594  & 0.025373  &  0.0051834 \\
$\bar{B}_s^0$         & 0.097318 &  0.035851  & 0.010628 & 0.0021711 \\
$\Lambda_b^0$       & 0.11664 &   0.042969  & 0.012738  &  0.0026021\\
$\Xi_b^{0,-}$ & 0.061520 &  0.022664  &  0.0067183  & 0.0013725 \\
$\Omega_b^-$  &  0.0031317 & 0.0011537  & 0.00034199  &  0.000069866 \\
$B_c^-$    &  0.010001 &  0.0031222  & 0.00080467  & 0.00012052 \\

\noalign{\smallskip}\hline\noalign{\smallskip}
$B_c^-/B^-$       & 0.043056  & 0.036489 & 0.031724 & 0.02326  \\
$B_c^-/b\bar{b}$       & 0.013276  & 0.011275 &	0.009817 &	0.00722  \\

\noalign{\smallskip}\hline
\end{tabular}
\end{center}
\caption{The statistical production yields per unit rapidity at mid-rapidity (higher part) of ground state open $b$ hadrons and $B_c^-$ mesons, alongside the production yield ratios (lower part) of $B_c^-$ to $B^-$ and to the total $b\bar{b}$ multiplicity ({\it i.e.} $B_c^-$'s production fraction $f_c$) in different centralities of $\sqrt{s_{\rm NN}}=5.02$\,TeV Pb-Pb collisions.}
\label{tab_Bc-yields-fractions}
\end{table*}

To perform realistic calculations for $\sqrt{s_{\rm NN}}=5.02$\,TeV Pb-Pb collisions, we focus on a fireball corresponding to a rapidity slice of one unit at mid-rapidity with vanishing net $c$ and $b$ number ($C=B=0$). We first determine the input $c$- and $b$-quark multiplicities $dN_{c\bar{c}}/dy$ and $dN_{b\bar{b}}/dy$ using the cross section $d\sigma_{c\bar{c}}/dy=1.165~{\rm mb}$~\cite{ALICE:2021dhb} and $d\sigma_{b\bar{b}}/dy=39.3~{\rm\mu b}$~\cite{He:2022tod,ALICE:2021mgk} in $\sqrt{s}=5.02$\,TeV $pp$ collisions at mid-rapidity and the thickness functions ($\langle T_{\rm AA}\rangle$) for different centralities~\cite{ALICE:2018tvk}. We have also applied 10-20\% reduction to $c$ quark multiplicity because of nuclear shadowing effect~\cite{He:2021zej,ALICE:2015sru}. For the input open $c$- and $b$-hadron spectrum, we use the listings by the particle data group (PDG)~\cite{ParticleDataGroup:2022pth}, augmented by additional states that have been predicted by relativistic quark models (RQM)~\cite{Ebert:2009ua,Ebert:2011kk} but not yet measured. The RQM predictions in particular feature many additional $c$ and $b$ baryons that are ``missing" in the current PDG listings~\cite{Ebert:2011kk}. The inclusion of these ``missing" states has proved pivotal for the successful reproduction of $c$ and $b$ baryon-to-meson ratios in $pp$ collisions via SHM calculations~\cite{He:2022tod,Dai:2024vjy,He:2019tik} that demonstrate significant enhancements relative to $e^+e^-$ collisions. For $B_c$ mesons, we have included all 14 states below the open $BD$ threshold~\cite{Eichten:2019gig}. The fireball volume is obtained by scaling the one $V_{\Delta y=1}=4997~{\rm fm^3}$ determined from SHM for light hadrons in the most central 0-10\% centrality~\cite{Andronic:2021erx} to other centralities using the measured charged-particle multiplicities~\cite{ALICE:2015juo}. Finally, the hadronization temperature is taken to be $T_H=170$\,TeV, which is higher than the pseudocritical chiral transition temperature ($T_{\chi}\sim 155$\,MeV~\cite{Borsanyi:2013bia,HotQCD:2014kol}) but seems more appropriate for the hadronization (confinement) transition of {\it heavy} hadrons~\cite{He:2022tod,He:2019tik,Becattini:2010sk,Bellwied:2013cta}. We've also checked that lowering down the hadronization temperature by $\sim 10$\,MeV does not cause significant change for the $b$-hadron production fractions~\cite{He:2022tod}.

The computed $c$ and $b$ fugacities for different centralities, alongside the corresponding fireball volumes and $c$- and $b$-quark multiplicities, are summarized in Table~\ref{tab_gamma_c-b}. The $\gamma_c$ exhibits a mild decrease from central to semicentral collisions, which then turns into an increase toward peripheral collisions, similar to the system size dependence of $\gamma_c$ found in~\cite{Andronic:2021erx}. In contrast, the $\gamma_b$ increases monotonously from central to peripheral centrality bins. Once the primary multiplicities from statistical production are computed, the total production yields of ground state open $b$ hadrons and $B_c^-$ mesons are obtained from the sum of the direct one and the feeddown contributions from excited states
\begin{align}\label{total-ground-states-multiplicity}
\langle N_{\alpha}^{\rm tot}\rangle=\langle N_{\alpha}\rangle + \sum_j\langle N_j\rangle\cdot \mathcal{B}(j\rightarrow\alpha),
\end{align}
where the branching fractions $\mathcal{B}$ for the strong decays of excited open $b$ hadrons have been estimated from a $^3P_0$ model~\cite{He:2022tod} and those for the strong or radiative decays of excited $B_c$ mesons to the ground state $B_c^-$ are all taken to be 100\%~\cite{Eichten:1994gt}. The resulting yields per unit rapidity of ground state $B^-$, $B_c^-$ and other open $b$ mesons and baryons, alongside the production yield ratios of $B_c^-$ to $B^-$ and to the total $b\bar{b}$ multiplicity ({\it i.e.} $B_c^-$'s production fraction $f_c$) in different centralities, are displayed in Table~\ref{tab_Bc-yields-fractions}. While the production ratios between open $b$ hadrons ({\it e.g.,} $\Lambda_b^0/B^-\sim0.5$) previously computed from grand-canonical SHM~\cite{He:2022tod} are well reproduced (all open $b$ hadrons containing a single $b$-quark suffer from the common canonical suppression not affecting their ratios), the $B_c^-/B^-$ reaches $\sim 0.043$ in central Pb-Pb collisions, amounting to a factor of $\sim 5$ enhancement with respect to the value in $pp$ collisions (cf. Sec.~\ref{sec_Bc-pp}). Accordingly the $B_c^-$'s production fraction $f_c\sim0.013$ in central collisions also represents a significant enhancement of the similar magnitude relative to $pp$ collisions, which is the prominent consequence of recombination production (implemented in a statistical fashion here) of $B_c$ mesons from abundant and highly thermalized $c$ and $b$ quarks in the QGP.

We also note that, while in central and semicentral collisions the exact conservation of $c$ number does not add to the canonical suppression for the production of $B_c$ mesons, {\it strict} conservation of $c$ number becomes important toward peripheral (especially 60-80\%) collisions where the $dN_{c\bar{c}}/dy$ reduces to the order of unity (canonical limit). We've numerical checked that, should only the $b$ number conservation be implemented in the statistical partition function (cf. Eq.~(\ref{Z-CB})), the $B_c^-$ yield in the 60-80\% centrality would be $\sim 70\%$ greater than the one shown in Table~\ref{tab_Bc-yields-fractions}. This mechanism of canonical suppression owing to {\it strict} conservation of the $c$ number largely explains the decrease of $B_c^-$'s production fraction toward peripheral collisions as indicated in Table~\ref{tab_Bc-yields-fractions}.

\section{$B_c$ $p_T$ spectrum \& nuclear modification factor in Pb-Pb collisions}
\label{sec_Bc-pT-spectrum-RAA}
We now convert the computed $B_c^-$'s statistical production yield into a $p_T$ distribution. Our strategy is to distribute the SHM yield according to the shape of the $p_T$ spectrum of $B_c^-$ calculated from the resonance recombination~\cite{He:2019vgs,He:2021zej,Ravagli:2007xx} of realistically transported $c$ and $b$ quarks. As an illustration, we focus on the 20-40\% centrality which is used as a proxy for the minimum bias (0-90\%) Pb-Pb collisions for which the CMS analysis was performed~\cite{CMS:2022sxl}.

While recombination as a hadronization mechanism for $c$ or $b$ quarks in the QGP dominates at low $p_T$, it yields to vacuum-like fragmentation at high $p_T$~\cite{He:2022tod,He:2019vgs}. It has been determined that more than 90\% of $b$ quarks diffusing in the QGP in central and semicentral Pb-Pb collisions undergo recombination at hadronization~\cite{He:2022tod}. Therefore, the {\it absolute} $p_T$ differential yield for the $B_c^-$'s recombination component is obtained by normalizing the $p_T$ spectrum of $B_c^-$ calculated from the resonance recombination of $c$ and $b$ quarks to its SHM yield corrected for the $b$-quark integrated recombination probability which is $\sim 92\%$ for the 20-40\% centrality~\cite{He:2022tod}. The resonance recombination model (RRM)~\cite{He:2019vgs,He:2021zej,Ravagli:2007xx} adopted here conserves 4-momentum and satisfies correct equilibrium limit~\cite{He:2019vgs}. The momentum distribution given by RRM for the recombination $B_c^-$ reads
\begin{align}\label{rrm}
f_{B_c}(\vec x,\vec p)&=C_{B_c} \frac{E_{B_c}(\vec p)}{m_{B_c}\Gamma_{B_c}}
\int\frac{d^3\vec p_1 d^3\vec p_2}{(2\pi)^3}f_b(\vec x,\vec p_1)f_{\bar c}(\vec x,\vec p_2)\nonumber\\
&\times \sigma_{B_c}(s)v_{\rm rel}(\vec p_1,\vec p_2)\delta^3(\vec p -\vec p_1-\vec p_2)\,
\end{align}
where $f_b$ and $f_{\bar{c}}$ are the transported phase space distributions of $b$ and $\bar{c}$ quarks, and $v_{\rm rel}$ is their relative velocity. In the current $2\rightarrow1$ formulation, recombination proceeds via a resonance cross section $\sigma_{B_c}(s)$ for $b+\bar{c}\rightarrow B_c^-$, taken to be of the Breit-Wigner form with the vacuum $B_c^-$ mass and a width $\Gamma_{B_c}\simeq 100$\,MeV (variations by a factor of $\sim 2$ have practically no effect on the shape of the $B_c^-$'s $p_T$ spectrum), while the $C_{B_c}$ ensures normalization to the statistical production yield. The phase space distributions $f_b$ and $f_{\bar{c}}$ are constructed on the hadronization hypersurface from Langevin simulations of $b$- and $c$-quark diffusion in the QGP down to $T_H=170$\,MeV and have been constrained by open $b$ and $c$ hadrons' observables~\cite{He:2022tod,He:2019vgs}. Since the RRM has been carried out here with the full quark phase space distributions, space-momentum correlations between $b$ and $\bar{c}$ quarks built up from Langevin simulations have been incorporated, which help produce a significantly harder meson spectrum compared to recombination only in the momentum space~\cite{He:2019vgs,He:2021zej}.

The high $p_T$ $B_c$ mesons are dominantly produced from fragmentation of $b$ quarks (the fragmentation probabilities for $\bar{c}\rightarrow B_c^-$ is two orders of magnitude smaller than those for $b\rightarrow B_c^-$~\cite{Braaten:1993jn}). To calculate the {\it absolute} $p_T$ differential yield for the $B_c^-$'s fragmentation component, we take the fragmenting $b$-quark spectrum after Langevin diffusion ({\it i.e.,} the $b$ quarks  left over from recombination, whose integrated number accounts for $\sim 8\%$ of total $b$ quarks~\cite{He:2022tod}), and simulate its fragmentation into $B_c^-$ using the fragmentation function~\cite{Braaten:1993jn}
\begin{align}\label{Braaten-Bc-FF}
	&D_{b\to B_c^-}(z) = N \frac{rz(1-z)^2}{[1-(1-r)z]^6} [ 6-18(1-2r)z \nonumber\\
	&+(21-74r+68r^2)z^2
	-2(1-r)(6-19r+18r^2)z^3 \nonumber\\
	&+3(1-r)^2(1-2r+2r^2)z^4],
\end{align}
where $z=p_T/p_t$ is the fraction of $B_c^-$'s $p_T$ relative to the parent $b$-quark's transverse momentum $p_t$. In Eq.~(\ref{Braaten-Bc-FF}), the parameter $r$ controlling the slope of the resulting spectrum is tuned such that the latter resembles the $B_c^-$'s $p_T$ spectrum in $pp$ collisions as much as possible especially at high $p_T$, while the normalization constant $N$ is tuned to ensure that the resulting $B_c^-$'s integrated yield should account for the same fraction ($f_c=0.00258\pm0.00062$, cf. Sec.~\ref{sec_Bc-pp}) of the total fragmenting $b$ quarks as in $pp$ collisions.

\begin{figure} [!tb]
\vspace{-0.5cm}
\includegraphics[width=1.05\columnwidth]{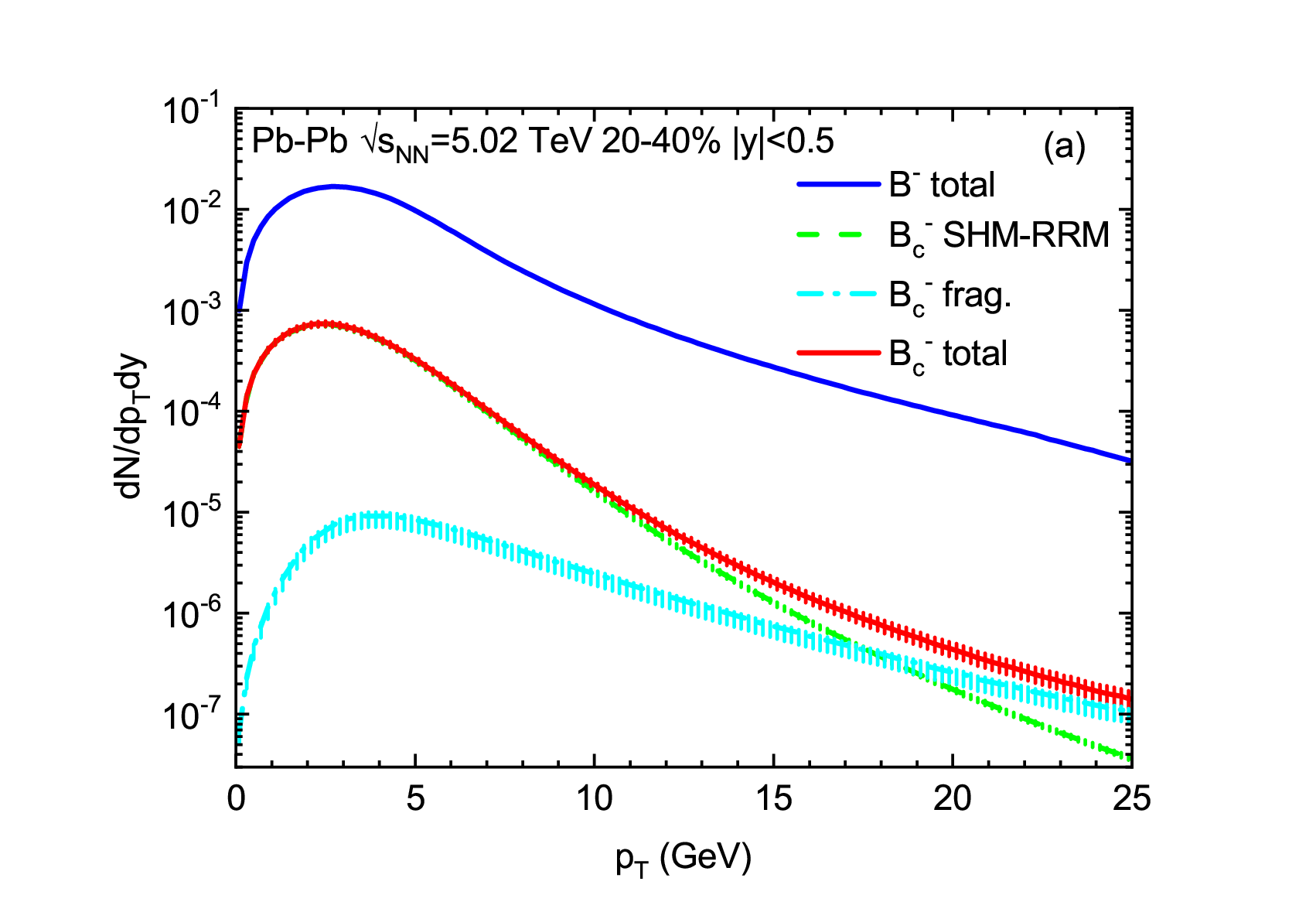}
\vspace{-0.5cm}
\includegraphics[width=1.05\columnwidth]{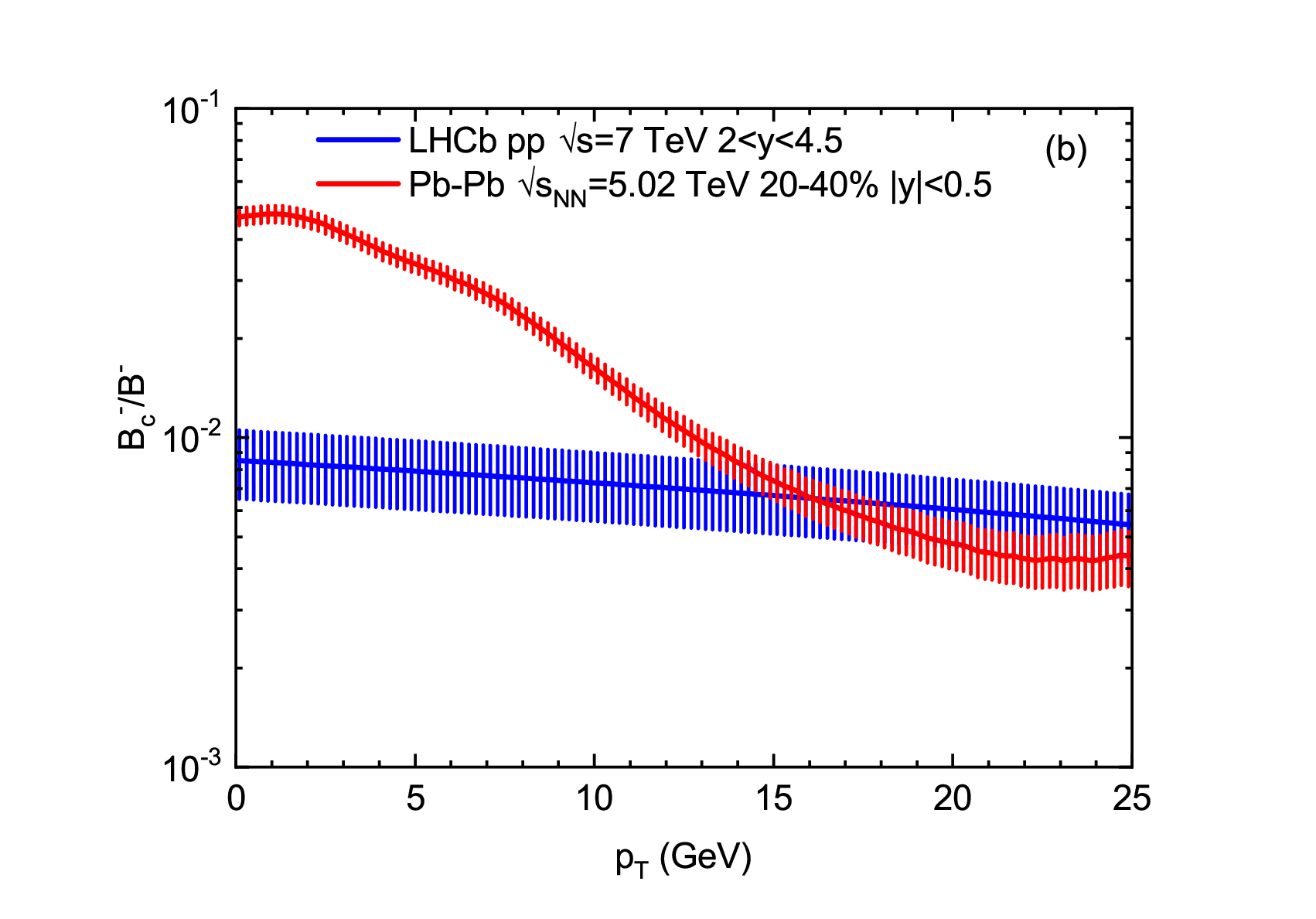}
\vspace{-0.5cm}
\includegraphics[width=1.05\columnwidth]{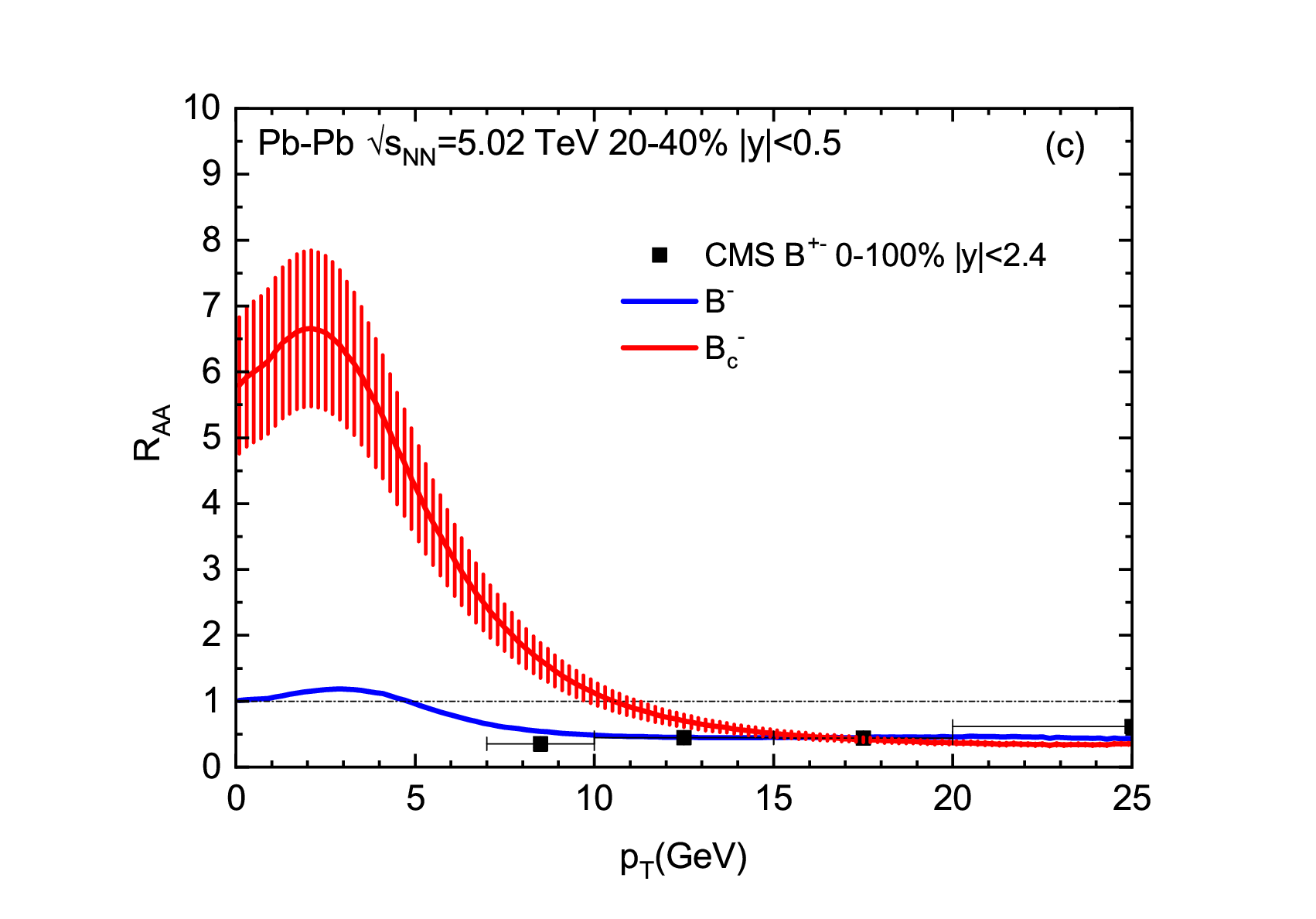}
\vspace{-0.5cm}
\caption{(a) The {\it absolute} $p_T$ differential yield for the recombination (green) and fragmentation (cyan) components of $B_c^-$, alongside those for the total $B_c^-$ (red) and $B^-$ (quoted from~\cite{He:2022tod}, blue) in the 20-40\% centrality of $\sqrt{s_{\rm NN}}=5.02$\,TeV Pb-Pb collisions. (b) $B_c^-/B^-$ as a function of $p_T$ in Pb-Pb (red) and $pp$ (\cite{LHCb:2019tea}, blue) collisions. (c) The nuclear modification factor of $B_c^-$ (red) and that of $B^-$ (quoted from~\cite{He:2022tod}, blue) in comparison with CMS data~\cite{CMS:2017uoy}. See text for the meaning of the widths of uncertainty bands.}
\label{fig_Bc-vs-B-PbPb}
\end{figure}

The resulting $p_T$ differential yield $dN/dp_Tdy$ for the recombination and fragmentation components, alongside their sum ({\it i.e.,} the total one), are shown in Fig.~\ref{fig_Bc-vs-B-PbPb}(a). The widths of bands indicate uncertainties from the spread in $f_c$ (or equivalently $\mathcal{B}(B_c^-\rightarrow J/\psi\mu^-\bar{\nu})$, cf. Sec.~\ref{sec_Bc-pp}) in $pp$ collisions (for the fragmentation component) or from the 10-20\% reduction of $c$-quark multiplicity due to the shadowing effect (for the recombination component, cf. Sec.~\ref{sec_Bc-SHM}). An immediate observation is that the momentum reach of $B_c^-$'s recombination extends to rather high $p_T\sim18$\,GeV, beyond which the fragmentation takes over. In Fig.~\ref{fig_Bc-vs-B-PbPb}(a), we have also plotted the $B^-$'s $p_T$ differential yield for comparison which was previously computed from the $b$-quark transport approach~\cite{He:2022tod} and is reproduced here; the integrated $B^-$ yield agrees with the value shown in Table~\ref{tab_Bc-yields-fractions}. The total $p_T$ differential yields for $B_c^-$ and $B^-$ shown in Fig.~\ref{fig_Bc-vs-B-PbPb}(a) are converted into their $p_T$ dependent ratio $B_c^-/B^-$ which is shown in Fig.~\ref{fig_Bc-vs-B-PbPb}(b) and compared to its counterpart in $pp$ collisions~\cite{LHCb:2019tea} (cf. Sec.~\ref{sec_Bc-pp}). The $B_c^-/B^-$ is enhanced by a factor of $\sim 5$ at low $p_T$ relative to $pp$ collisions, but tends to the value in the latter at very high $p_T$ within uncertainties. Accordingly, the $B_c^-$'s nuclear modification factor defined as
\begin{align}
R_{\rm AA}(p_T)=\frac{dN^{\rm PbPb}/dp_Tdy}{\langle T_{\rm AA}\rangle d\sigma^{pp}/dp_Tdy}
\end{align}
reaches the value of $\sim 5$-6 at $p_T<5$\,GeV, in contrast to the $B^-$'s $R_{\rm AA}$ (quoted from~\cite{He:2022tod}) around unity in the same $p_T$ range, as shown in Fig.~\ref{fig_Bc-vs-B-PbPb}(c). The enhancement of $B_c^-$'s production in Pb-Pb collisions as characterized by its $R_{\rm AA}$ above unity persists till $p_T\sim 10$\,GeV, at which the $B^-$'s production is already suppressed by a factor of $\sim 2$ ($R_{\rm AA}\simeq 0.48$) relative to $pp$ collisions. The width of the band for $B_c^-$'s $R_{\rm AA}$ at low $p_T$ indicates uncertainties due to shadowing of $c$'s participating in the statistical recombination as well as the spread of the $pp$ reference spectrum, but is dominated by the latter. At high $p_T$ where the fragmentation component dominates, the uncertainties from the spread in $f_c$ (or $\mathcal{B}(B_c^-\rightarrow J/\psi\mu^-\bar{\nu})$) in the numerator spectrum and the denominator reference cancel, making the band become narrower.

\begin{figure} [!tb]
\vspace{-0.5cm}
\includegraphics[width=1.05\columnwidth]{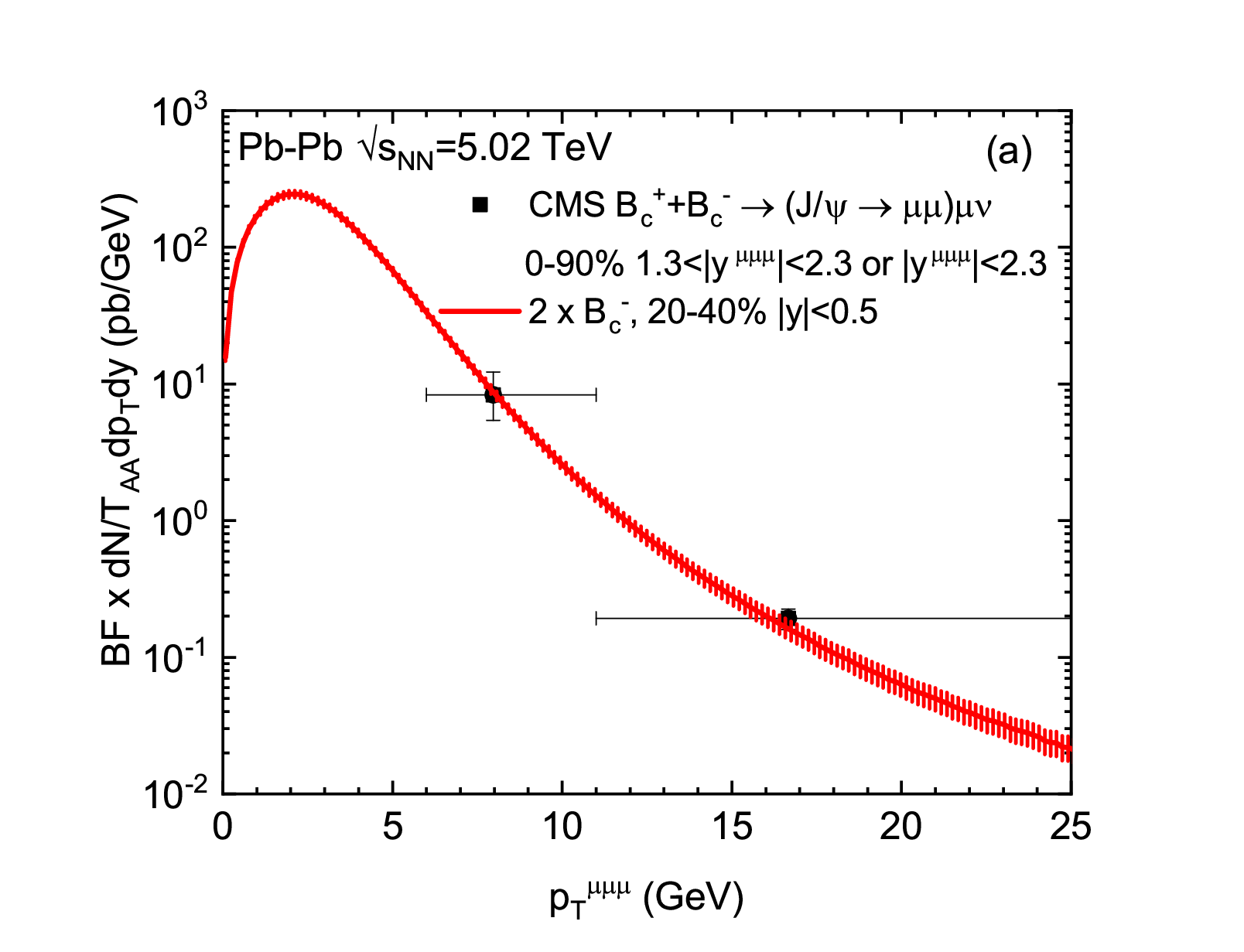}
\vspace{-0.5cm}
\includegraphics[width=1.05\columnwidth]{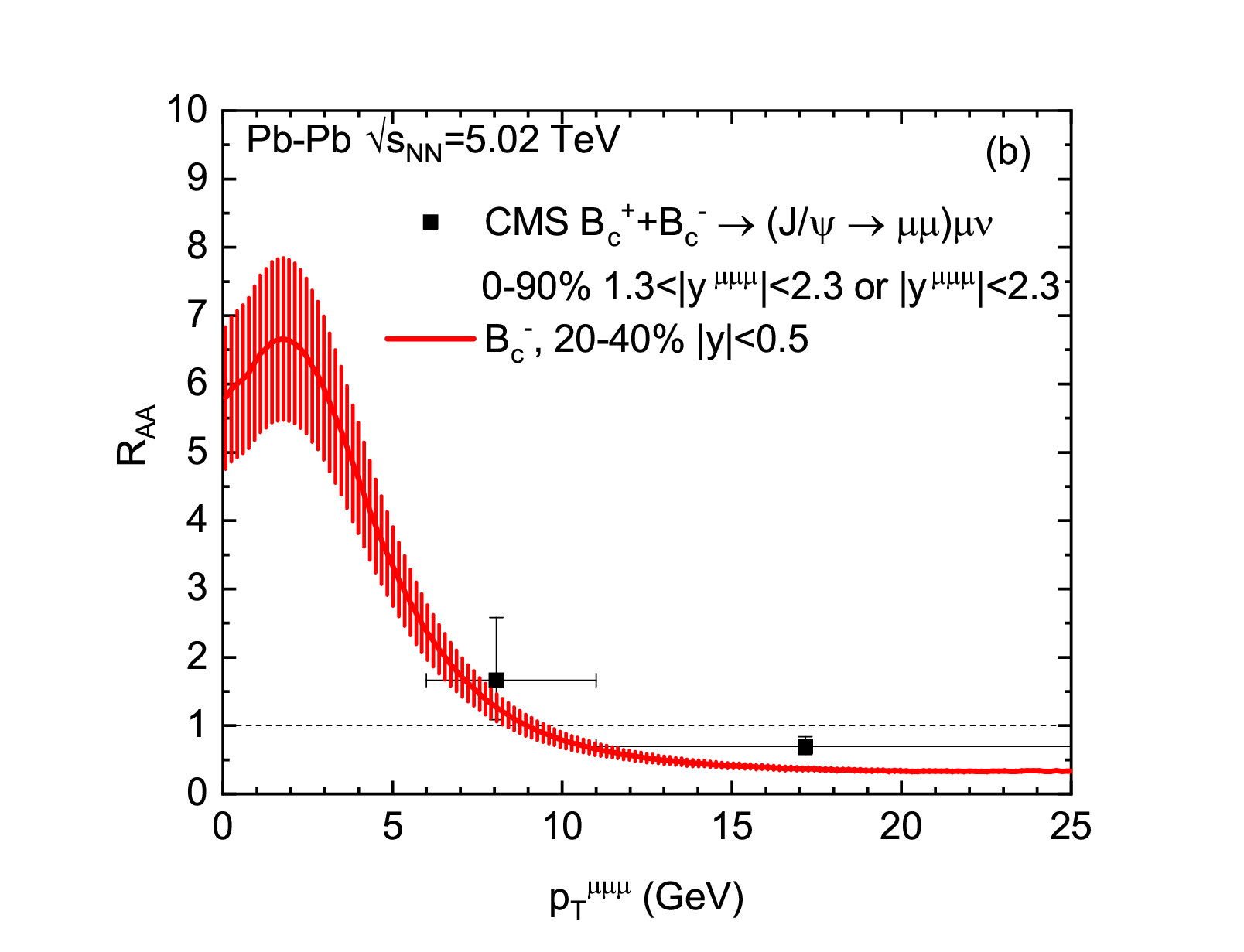}
\vspace{-0.5cm}
\caption{ The $B_c^-$'s (a) normalized $p_T$ differential yield and (b) nuclear modification factor as a function of three-muon's transverse momentum $p_T^{\mu\mu\mu}$, compared to CMS data~\cite{CMS:2022sxl}. The widths of the uncertainty bands have the same meaning as for Fig.~\ref{fig_Bc-vs-B-PbPb}.}
\label{fig_Bc-vs-pTmumumu}
\end{figure}

Same as for the $p_T$ differential cross section of $B_c$ mesons in $pp$ collisions (cf.~Fig.~\ref{fig_Bc-pT-spectrum-pp}(b)), the CMS data~\cite{CMS:2022sxl} for their $p_T$ differential yield in minimum bias (0-90\%) Pb-Pb collisions were presented for $B_c^-+B_c^+$, multiplied by the pertinent branching fractions of the two-step decay process $B_c\rightarrow (J/\psi\rightarrow\mu\mu)\mu\nu$ and normalized by the thickness function, in two bins of the three-muons' transverse momentum $p_T^{\mu\mu\mu}$. To compare to the CMS data, the $dN/dp_Tdy$ for $B_c^-$ shown in Fig.~\ref{fig_Bc-vs-B-PbPb}(a) is first doubled, then multiplied by $\mathcal{B}(B_c^-\rightarrow J/\psi\mu^-\bar{\nu})=1.95\%\pm 0.46\%$~\cite{LHCb:2019tea} and $\mathcal{B}(J/\psi\rightarrow\mu\mu)=5.96\%$~\cite{ParticleDataGroup:2022pth}, divided by the corresponding $\langle T_{\rm AA}\rangle$~\cite{ALICE:2018tvk}, and re-plotted as a function of $p_T^{\mu\mu\mu}=0.85\cdot p_T^{B_c}$~\cite{CMS:2022sxl} (cf. Sec.~\ref{sec_Bc-pp}) in Fig.~\ref{fig_Bc-vs-pTmumumu}(a). The two data points measured by CMS are fairly described within uncertainties. This normalized $p_T$ differential yield is then divided by the $p_T$ differential cross section in $pp$ collisions shown in Fig.~\ref{fig_Bc-pT-spectrum-pp}(b) to obtain the $B_c$'s $R_{\rm AA}$ as a function of $p_T^{\mu\mu\mu}$, which is shown in Fig.~\ref{fig_Bc-vs-pTmumumu}(b) and compared to the corresponding CMS data. While the data point in the lower $p_T$ bin is fairly reproduced within theoretical and experimental uncertainties, the suppression indicated in the higher $p_T$ bin is somewhat overestimated by our calculation.

\begin{figure} [!tb]
\vspace{-0.5cm}
\includegraphics[width=1.05\columnwidth]{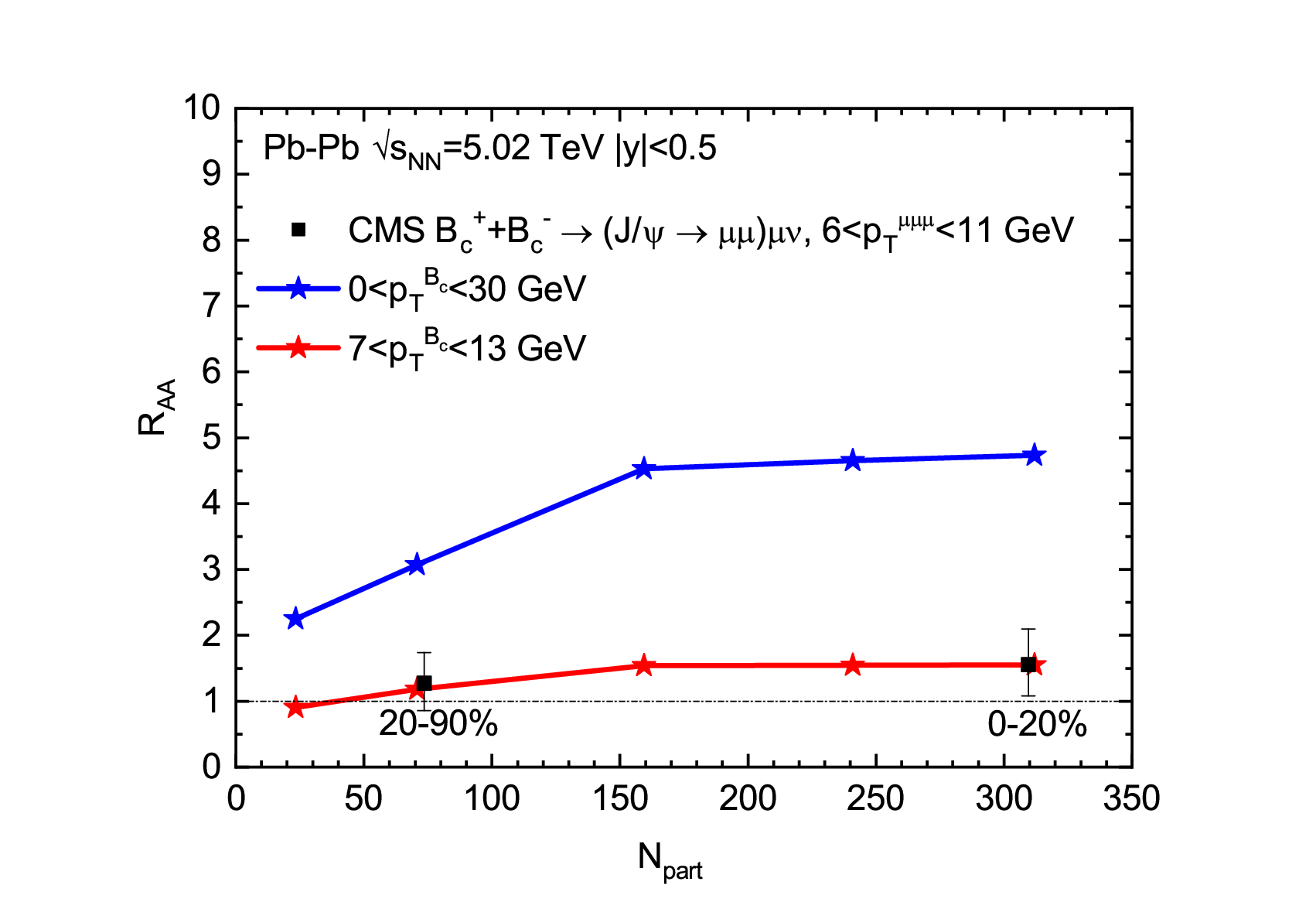}
\vspace{-0.5cm}
\caption{The $B_c^-$'s integrated nuclear modification factor in the full $p_T$ range vs. in the $7<p_T<13$\,GeV interval. The latter corresponds to $6<p_T^{\mu\mu\mu}<11$\,GeV in CMS data~\cite{CMS:2022sxl}.}
\label{fig_integrated_RAA}
\end{figure}

Finally we calculate $B_c$'s integrated $R_{\rm AA}$ in the full $p_T$ range and in the $7<p_T<13$\,GeV interval (corresponding to $6<p_T^{\mu\mu\mu}<11$\,GeV in the CMS data~\cite{CMS:2022sxl}) as a function of participant numbers. As shown in Fig.~\ref{fig_integrated_RAA}, the integrated $R_{\rm AA}$ in the full $p_T$ range reaches $\sim 5$ in central and semicentral collisions but gradually drops off toward peripheral collisions, in line with the centrality dependence of $B_c^-$'s statistical production fraction shown in Table~\ref{tab_Bc-yields-fractions}. The enhancement embodied in the calculated integrated $R_{\rm AA}$ in the higher $p_T$ interval is much milder, amounting to a factor of $\sim 50\%$ in central collisions and vanishing in peripheral collisions, fully consistent with the CMS data points in two centrality bins within uncertainties.

\section{Sumary}
\label{sec_summary}

In this work, we have investigated the recombination production of $B_c$ mesons in Pb-Pb collisions at the LHC energy using a statistical hadronization approach. By treating $B_c^-$ as a member of the family of open $b$ hadrons that contain a single $b$-quark and implementing the {\it strict} conservation of $b$ and $c$ numbers, we were able to make quantitative predictions for $B_c^-$'s production fraction relative to the total $b{\bar b}$ multiplicity and its relative production to $B^-$ mesons in the presence of QGP, both demonstrating a factor of $\sim 5$ (3) enhancement  in central (peripheral) Pb-Pb collisions with respect to $pp$ collisions. The statistical production yield of $B_c^-$ mesons is converted into a $p_T$ differential distribution according to the shape of $B_c^-$'s $p_T$ spectrum computed from resonance recombination of realistically transported $b$ and $\bar{c}$ quarks in the QGP. This was supplemented with the fragmentation of high $p_T$ $b$ quarks and enabled us to compute the $p_T$ dependent nuclear modification factors for $B_c^-$ mesons that reach $\sim$5-6 at low $p_T$ in central collisions, with the reference spectrum carefully constructed from previously determined differential cross section of $B^-$ mesons and the experimentally measured $B_c^-/B^-$ ratio. Pertinent data measured by the CMS experiment, including $B_c^-$'s normalized $p_T$ differential yield, nuclear modification factors in two $p_T$ intervals in the minimum bias collisions and the integrated ones in two centrality bins, can all be fairly described in our approach within theoretical and experimental uncertainties. Our study provides strong support for the formation of $B_c$ mesons through (statistical) recombination of abundant and highly thermalized $c$ and $b$ quarks in the deconfined QGP created in Pb-Pb collisions at the LHC energy.\\

{\bf Acknowledgments:}
This work was supported by the National Natural Science Foundation of China (NSFC) under Grant No.12075122.

\end{document}